# Direct observation of strong anomalous Hall effect and proximity-induced ferromagnetic state in SrIrO$_3$

*Arun Kumar Jaiswal, Di Wang, Vanessa Wollersen, Rudolf Schneider, Matthieu Le Tacon, Dirk Fuchs\**

A. K. Jaiswal, R. Schneider, M. Le Tacon, D. Fuchs

Karlsruhe Institute of Technology, Institute for Quantum Materials and Technologies, 76021 Karlsruhe, Germany;

D. Wang, V. Wollersen

Karlsruhe Institute of Technology, Institute of Nanotechnology and Karlsruhe Nano Micro Facility, 76021 Karlsruhe, Germany;

E-mail: dirk.fuchs@kit.edu

The 5$d$ iridium-based transition metal oxides have gained broad interest because of their strong spin-orbit coupling which favors new or exotic quantum electronic states. On the other hand, they rarely exhibit more mainstream orders like ferromagnetism due to generally weak electron-electron correlation strength. Here, we show a proximity-induced ferromagnetic (FM) state with $T_C \approx 100$ K and strong magnetocrystalline anisotropy in a SrIrO$_3$ (SIO) heterostructure via interfacial charge transfer by using a ferromagnetic insulator in contact with SIO. Electrical transport allows to selectively probe the FM state of the SIO layer and the direct observation of a strong, intrinsic and positive anomalous Hall effect (AHE). For $T \leq 20$ K, the AHE displays unusually large coercive and saturation field, a fingerprint of a strong pseudospin-lattice coupling. A Hall angle, $\sigma_{xy}^{AHE}/\sigma_{xx}$, larger by an order of magnitude than in typical 3$d$ metals and a FM net moment of about 0.1 $\mu_B$/Ir, is reported. This emphasizes how efficiently the nontrivial topological band properties of SIO can be manipulated by structural modifications and the exchange interaction with 3$d$ TMOs.

## 1. Introduction

The search for new materials for next generation information technology has strongly pushed scientific work in the field of spintronics, which addresses in particular the mutual influences of spin and charges on electronic transport. Spin-orbit coupling, which is naturally present in heavy metals, provides interesting perspectives for the manipulation of spin-transport and, in this respect, the 5$d$ iridium-based transition metal oxides (TMOs) of the Ruddlesden-Popper series $Sr_{n+1}Ir_nO_{3n+1}$ have gained a lot of interest. The iridates indeed display SOC which is on a similar energy scale than that of the electron-correlation or the electronic bandwidth,[1] which favors new or exotic quantum electronic states.[2–6] However, in contrast to archetypical correlated 3$d$ TMOs, the electron-electron correlation strength is often too small in the 5$d$ TMOs to host ferromagnetism.

For $Sr_2IrO_4$ ($n$ = 1), the SOC results in a spin-orbital mixed state of the $Ir^{4+}$ ion with a filled quadruplet pseudospin state $J_{eff}$ = 3/2 and a half-filled doublet $J_{eff}$ = 1/2.[7] Magnetic interaction of neighbored pseudospins leads to a basal ($ab$)-plane canted antiferromagnetic (AFM) Mott-insulator ground state with pseudospins locked to the oxygen octahedral rotation.[8–10] For $n$ = 2, interlayer coupling weakens which leads to a spin-flop transition of the pseudospins with out-of-plane spin alignment along the $c$-axis and $T_N$ = 280 K.[11] In contrast, the perovskite phase $SrIrO_3$ (SIO) ($n$ = ∞) displays paramagnetic semi-metallic behavior due to an increased hybridization of Ir5$d$ and O2$p$ orbitals.[3,12–15] Nevertheless, SIO is on the verge of a magnetic ground state and may display AFM or ferromagnetic (FM) properties as well, depending on the details of the Hubbard interaction $U$ and the SOC.[12] Owing to a strong pseudospin-lattice coupling,[16] these can be finely tuned by structural modifications, especially with respect to the network of the corner-sharing $IrO_6$ octahedra which in turn enables a manipulation of the magnetism in SIO.

The bulk structure of SIO consists in a distorted orthorhombic perovskite structure with in-phase and antiphase rotations of the $IrO_6$ octahedra ($a^-a^-c^+$ in Glazer notation).[17,18] However, a suppression of octahedral out-of-plane tilts, akin to the rotation pattern of $Sr_2IrO_4$ can be achieved when ultra-thin SIO films are epitaxially grown on cubic $SrTiO_3$ (STO) which concomitantly yields a metal-to-insulator transition (MIT).[19] Other type of structural distortions are likewise discussed as a possible source for magnetic properties of SIO.[20] For example, in SIO/STO superlattices the $IrO_6$ rotation pattern supports an AFM ground state,[21,22] where the ordering temperature $T_N$ can be controlled by the interlayer coupling, $i.~e.$, by the STO thickness[23] or epitaxial strain.[24]

Meanwhile a lot of activities have been focused on SIO-based heterostructures including magnetic active layers, which seems to be a promising route to design new systems exhibiting both, SOC and ferromagnetism. A topological Hall effect has been reported for $SIO/SrRuO_3$ heterostructures[25] demonstrating the ability of the strong SOC of SIO to influence the magnetic properties of an itinerant ferromagnet. In manganite/SIO superlattices magnetic exchange between the different layers may also result in interfacial FM properties.[26–34] These can in principle be probed through magnetotransport as effects such as the anomalous Hall effect (AHE) or the anisotropic magnetoresistance ($AMR$), two hallmarks of a FM metal, directly relate to $e.~g.$, the magnetization, ordering temperature and anisotropy of the magnetic state and are therefore useful quantities to characterize the magnetic properties of materials.

The main drawback of manganite/SIO heterostructures with respect to an analysis of the magnetic state of SIO is that the manganite layer, if in contact with SIO becomes conductive due to interfacial charge transfer, resulting in a distinct contribution to the AHE. The conductivity is well explained by the $e_g$-double exchange mechanism in electron- or hole-doped manganites.[35] In this work, we overcome this difficulty by presenting a detailed study of heterostructures composed of SIO and $LaCoO_3$ (LCO). Epitaxially strained LCO is a FM insulator with a $T_C$ ≈ 85 K. As shown in this work, light electron-doping of LCO due to interfacial charge transfer or others does not lead to any measurable conductivity which is very likely due to the spin-blockade phenomenon of cobaltites.[36] This ideally allows unambiguous selective characterization of the magnetism induced at the SIO/LCO interfaces by electronic transport. We observe an interfacial electron transfer from SIO to LCO and show that the heterostructures reveal a strong positive AHE and a four-fold symmetric *AMR*. This indicates the formation of a proximity-induced FM state in SIO with $T_C \approx 100$ K and a <110> in-plane magnetic easy-axis. Furthermore, the AHE displays unusually high coercivity and saturation field at low *T*, alongside a rather large Hall angle.

These results demonstrate how efficiently the nontrivial topological band properties of SIO can be tuned by structural modifications at correlated oxide interfaces, which provides new promising routes to functionalize these materials.

## 2. Results and Discussion

### 2.1 Structural properties of SrIrO$_3$/LaCoO$_3$ heterostructures

In the following, we will concentrate on three different types of heterostructures: (i) *SIO single-layer*, *i. e.*, 10 monolayers (ML) of SIO capped with a STO (4 ML) protection layer, (ii) *LCO/SIO bilayer*, *i. e.*, 10 ML of LCO on 10 ML of SIO, and (iii) *LCO/STO/SIO trilayer*, where the LCO and SIO layer are separated by 4 ML of epitaxial STO. The heterostructures were produced by pulsed laser deposition on TiO$_2$-terminated (001) STO, as described in the *Experimental Section/Methods*.[37] In **Figure 1**a, we report cross-sectional high-resolution scanning transmission electron microscopy (HR-STEM) images of our LCO/SIO bilayer documenting stoichiometric composition and atomically sharp LCO/SIO and SIO/STO interfaces with atomic interdiffusion over distances that do not exceed 1 ML.

The structural properties of the heterostructures were further analyzed by x-ray diffraction using a Bruker D8 Davinci diffractometer equipped with Cu K$\alpha$ radiation. The reciprocal space maps on pseudo-cubic {204} lattice reflections shown in **Figure 1**b clearly indicate a pseudomorphic growth of the SIO and LCO layer. The peak maxima of the films and the STO substrate appear at $h = 2$, *i. e.*, in-plane lattice parameters are identical to STO ($a = 3.905$ Å). Despite the line broadening of the film peaks along the *l*-direction, a distinct variation of the out-of-plane lattice spacing for the different azimuth angles is not observed. In contrast to thicker SIO films ($t \geq 17$ nm) for which we observed clear orthorhombic distortion,[38] the pseudomorphic growth of the thin SIO layer obviously results in a suppression of octahedral distortion. The pseudo-tetragonal structures of the compressed SIO film results in in-plane lattice parameters $a = b = 3.9$ Å, and an out-of-plane lattice parameter of $c = 4.05$ Å. The tensile strained LCO film displays $a = b = 3.9$ Å and $c = 3.78$ Å, in contrast to the pseudo-cubic bulk value of $a = 3.83$ Å.[39]

To analyze the octahedral tilt pattern of the IrO$_6$ and CoO$_6$ octahedra in more detailed, we carried out measurements on specific half-integer pseudo-cubic lattice reflections. The occurrence of such half-integer reflections indicates a doubling of the pseudo-cubic unit-cell and allows to determine the octahedral tilt and rotation pattern of the distorted perovskite structure. Antiphase rotations (-) along the *a, b,* and *c*-axis produce half-integer (*hkl*) reflections odd-odd-odd with $k \neq l$, $h \neq l$, and $h \neq k$, respectively, whereas in-phase rotations (+) along the corresponding axes produce reflections even-odd-odd ($k \neq l$), odd-even-odd ($h \neq l$), and odd-odd-even ($h \neq k$).[17] For the heterostructure we did not observe any reflections indicating in-phase rotations. Antiphase-rotations are documented only for SIO along the *c*-axis, see **Figure 1**c (left), which results in a one-tilt rotation pattern of $a^0a^0c^-$, hinting to a *I4/mcm* (No. 140) symmetry. The same behavior is also observed for a 10 ML thick SIO single layer.

Intensities and peak positions of the SIO half-integer reflections obtained from the LCO/SIO bilayer and the SIO single layer are nearly identical for both (see Figure S3b), demonstrating same octahedral distortion and epitaxial strain for both samples.

The suppression of out-of-plane octahedral tilts along the *a*- and *b*-axis in SIO films grown with a thickness of only a few monolayers on STO was also reported by other groups[19,40] and documents not only lattice- but also bond-angle adaption of SIO due to epitaxy. Depending on the film growth and epitaxial strain, structural relaxation towards the orthorhombic bulk structure also results in out-of-plane rotations along the *a*- and *b*-axes and therefore in the occurrence of corresponding half-integer reflections. For example, on the right side of Figure 1c we have shown the half-integer reflections of LCO films grown on STO for different film thickness. For the 10 ML thick film, peak intensity is too small to be detected, very likely due to the lower x-ray scattering amplitude of Co in comparison to Ir (see also Figure 1b). With increasing film thickness, peaks corresponding to the $a^-a^-c^-$ rotation pattern of bulk rhombohedral LCO appear.[41] Interestingly, the (1/2 1/2 3/2) peak intensity which indicates out-of-plane antiphase rotations increases much stronger compared to the in-plane anti-phase rotation along the *c*-axis. Since the peak intensity is correlated to the strength of the octahedral tilt, this strongly suggests stronger suppression of the out-of-plane rotations of the $CoO_6$ octahedra, like the situation for SIO on STO. We can therefore infer from this that a rotation pattern $a^0a^0c^-$ is occurring as well for the 10 ML thick LCO layer of our heterostructure. In **Figure 1**d we have sketched the octahedral rotation pattern of the $BO_6$ (B = Co, Ir, and Ti) octahedra of the bilayer heterostructure.

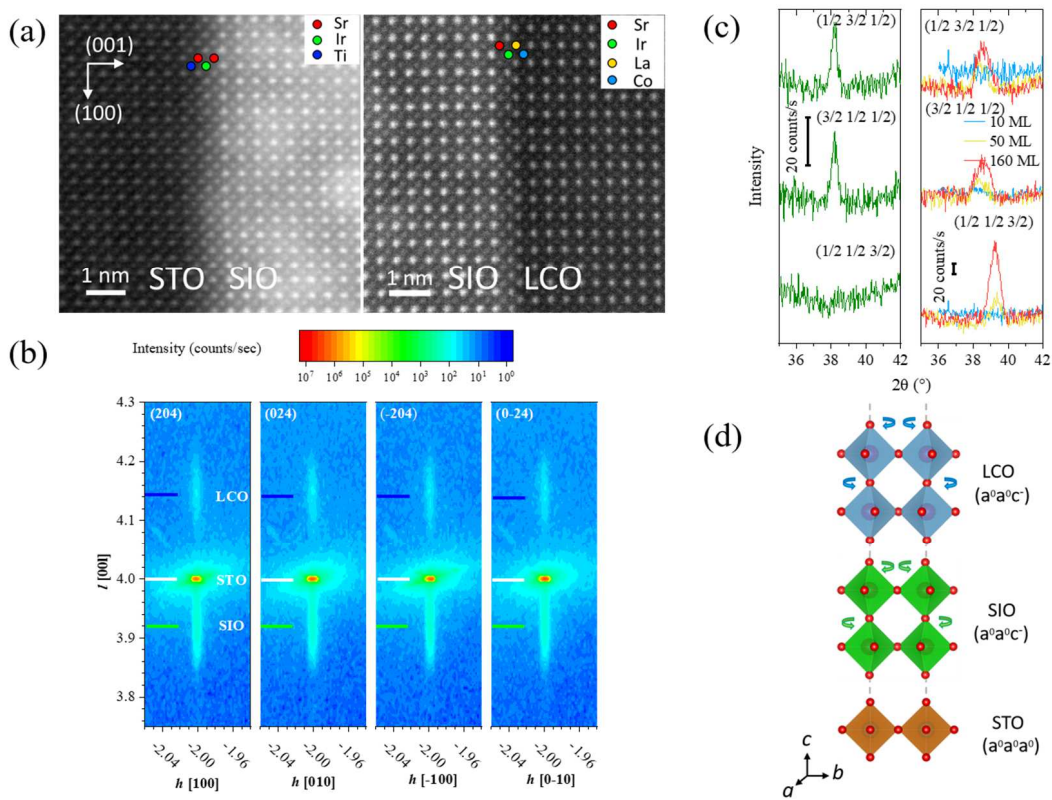

**Figure 1.** (a) HR-STEM micrograph showing the STO/SIO and the SIO/LCO interface of a LCO/SIO heterostructure. The interfaces are sharp and well defined. Elements are indicated by color. (b) Reciprocal space maps of the SIO/LCO heterostructure. The maps are recorded in the vicinity of the {204} STO reflection. The intensity is plotted as a function of the scattering

vector *q* expressed in non-integer Miller indices *h*, *k*, and *l* of the STO substrate reflection. LCO, STO and SIO film peaks are indicated and do not show distinct orthorhombic distortion. (c) θ/2θ scans on specific half-integer asymmetric pseudo-cubic lattice reflections on SIO of the heterostructure (left) and LCO films on STO (right). The presence or absence of the reflections indicate an $a^0a^0c^-$ octahedral rotation pattern for SIO and LCO of the heterostructure (see text). (d) Schematic of the octahedral rotation pattern of LCO and SIO of the heterostructure. The counterclockwise antiphase rotation along the out-of-plane c-axis of the $CoO_6$ and $IrO_6$ octahedra are indicated.

**2.2 Electronic transport and magnetism in SIO**

The electronic transport properties of the samples were determined by four-point resistance measurements on microbridges.[37] Single layers of LCO on STO and STO display insulating behavior. Even small electron or hole doping of LCO does not result in any measurable conductance (see supporting information[37]) so that the resulting conductivity of the LCO/SIO bilayer and LCO/STO/SIO trilayer is completely related to the SIO layer, see also below.

The temperature dependence of the longitudinal resistivity $\rho_{xx}$ of the three types of heterostructures is shown in **Figure 2**a. For the SIO film and the LCO/STO/SIO trilayer, $\rho_{xx}(T)$ is nearly the same and well comparable to previous reports,[15,42–44] attesting that the electronic properties of SIO are not impacted by LCO in the trilayer. In stark contrast, the LCO/SIO bilayer shows significant increase of resistivity and appears closer to a MIT. For SIO films thinner than 4ML grown on STO, a MIT has been reported[45] and interpreted as originating from a strain-induced suppression of the out-of-plane rotation of $IrO_6$ octahedra,[19] similar to that of AFM insulating $Sr_2IrO_4$.[46] However, since here the distinct increase of $\rho_{xx}$ is only observed for the bilayer, we attribute this primarily to an electron transfer from SIO to LCO, see below. In **Figure 2**b we have shown $\rho_{xx}(T)$ for a LCO/SIO* bilayer where the SIO thickness is only 6 ML thick in comparison to a 6 ML thick SIO single layer. The single layer shows increased resistivity with respect to the 10 ML thick single layer which indicates the approach to a metal-to-insulator transition. When LCO is deposited on top, the LCO/SIO* bilayer sample becomes insulating below about 130 K very likely due to interfacial charge transfer (see also section **2.4**). However, the charge transfer to LCO, which can be assumed to be the same as for the 10 ML thick LCO/SIO bilayer, does not lead to any measurable conductivity of the sample. Therefore, any contributions from LCO to the conductivity or AHE in the LCO/SIO bilayer can be definitely ruled out. For *T* > 100 K, $\rho_{xx}(T)$ of the LCO/SIO bilayer displays logarithmic *T*-dependence, as shown in the inset of Figure 2a, characteristic of magnetic spin-flip scattering ($\rho_{xx} \propto -\ln T$).[47] Its strong decrease below about 100 K hints to an ordering of magnetic moments. The suppression is quite significant and indicates magnetic ordering not only at the interface, *i. e.*, of the Co ions, but also of the Ir ions in the conducting SIO layer.

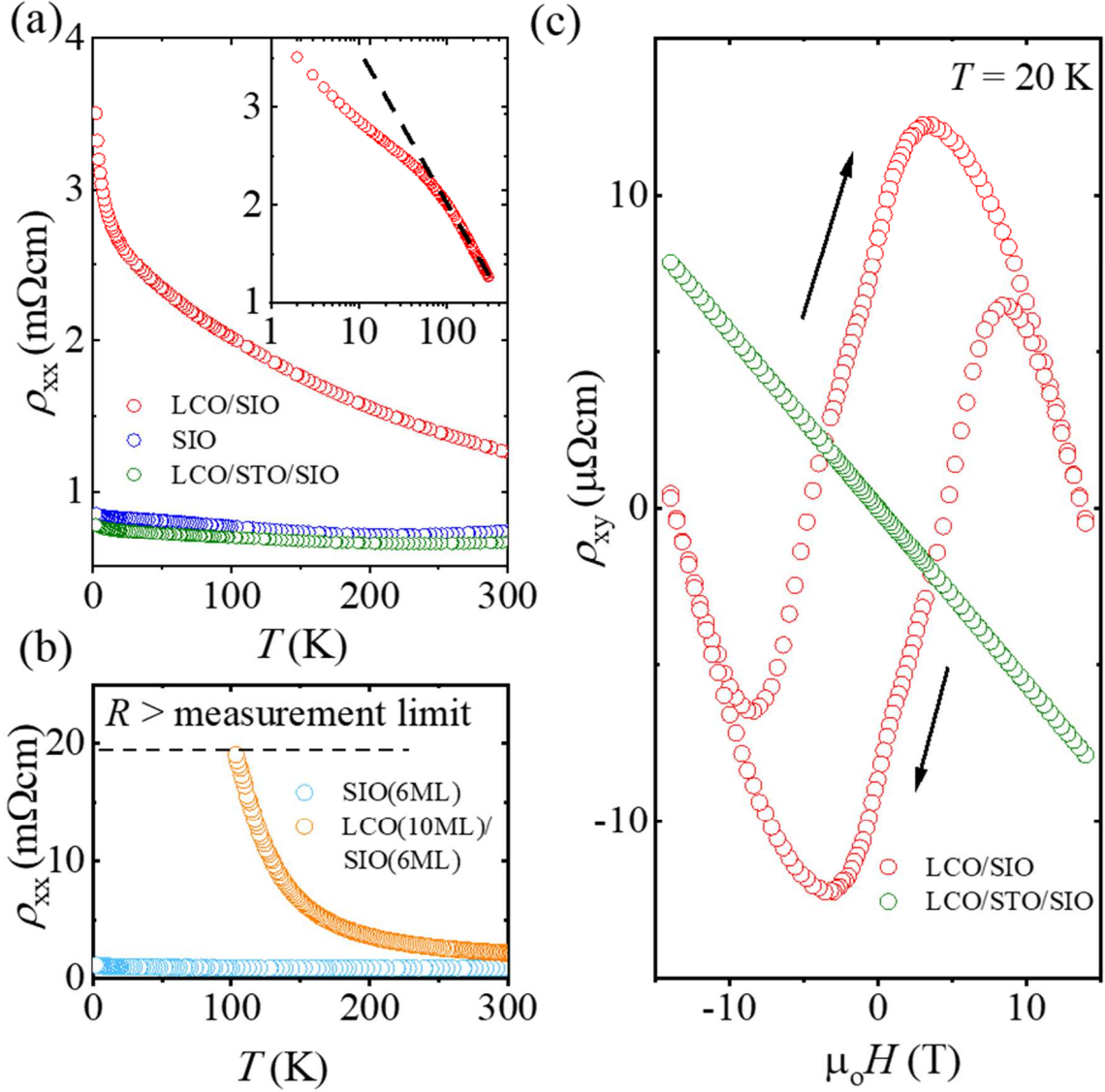

**Figure 2**. (a) Longitudinal resistivity $\rho_{xx}(T)$ of the SIO heterostructures. The inset displays $\rho_{xx}(T)$ of LCO/SIO on a semi-log scale to visualize the logarithmic $T$-dependence (dashed line in the inset). (b) $\rho_{xx}(T)$ of a 6 ML thick SIO single layer in comparison to a LCO/SIO* bilayer with same SIO thickness and 10 ML of LCO. The bilayer shows completely insulating behavior below 130 K. (c) Transverse resistivity $\rho_{xy}$ of the LCO/SIO bilayer and the LCO/STO/SIO trilayer at $T = 20$ K

In **Figure 2c**, the transverse Hall resistivity $\rho_{xy}$ versus $\mu_0 H$ of the LCO/SIO and LCO/STO/SIO is shown for $T = 20$ K. There again, electronic transport of SIO is comparable to that of the trilayer LCO/STO/SIO so that we only focus here on the comparison between the bilayer and the trilayer, in which the LCO and SIO layers are directly in contact or not, respectively. The trilayer displays electron-like linear behavior of $\rho_{xy}$ versus $\mu_0 H$ and indicates dominant one-type charge carrier ordinary Hall effect (OHE) in the measured field-range. In contrast, $\rho_{xy}$ of the bilayer is dominated by strong positive anomalous Hall effect (AHE), a clear indication for magnetism in SIO. Here, the Hall resistivity is expressed by $\rho_{xy} = \rho_{xy}^{OHE} + \rho_{xy}^{AHE}$. The linear field dependence of $\rho_{xy}$ at high fields indicates single type charge carrier transport and magnetic saturation. Therefore, the anomalous part of the Hall resistance has been obtained by subtraction of the linear part from $\rho_{xy}$. The OHE caused by Lorentz force, is linear and well described by single band electron transport alike, $\rho_{xy}^{OHE} = R^O \times \mu_0 H$, with the Hall constant $R^O$. In comparison to the trilayer, $R^O$ is obviously larger, indicating a smaller electron concentration of SIO in the bilayer.

Generally, the AHE is proportional to the magnetization, *i. e.*, $\rho_{xy}^{AHE} = R^A M$, with $R^A$ depending on specific material parameters and the longitudinal conductivity $\sigma_{xx}$.[48] In **Figure 3**a we show the anomalous Hall contribution $\rho_{xy}^{AHE}$ of the LCO/SIO bilayer for various temperatures. $\rho_{xy}^{AHE}$ displays hysteretic behavior with rather large coercive and saturation field below 50 K. For $T \leq 10$ K, the coercive field $H_c$ reaches up to 5 T and saturation is obviously not achieved even for $\mu_0 H = 14$ T. The hysteretic behavior is perfectly fitted by a modified Heaviside-step function where $M$ is given by: $M = M_s \times \tanh(h \times (H \pm H_c))$, with $M_s$, $h$ and $H_c$ being the saturation magnetization, the slope at $H_c$, and the coercive field, respectively. Fitting parameters are listed in Table S1, see supporting information. With increasing temperature, the saturation value of $\rho_{xy}^{AHE}$, <AHE>, decreases. The AHE displays small contributions from a second hysteresis loop which hints to a slightly inhomogeneous magnetic state of the SIO layer. **Figure 3**b shows <AHE> versus $T$ of LCO/SIO bilayers for 10 ML and 25 ML of SIO. The $T$-dependence of <AHE> demonstrates the onset of FM behavior at $T_C \approx 100$ K, which is close to the $T_C$ of LCO.

The charge transfer from SIO to LCO, as indicated above by the OHE (and which will be quantified more specifically in section **2.4**) results in small electron doping of LCO which is known to rather suppress the double exchange and $T_C$.[37,49] Also, as shown in Figure 2c, LCO in contact with SIO does not contribute to conductivity. Therefore, possible contributions from the interfacial LCO layer to $\rho_{xy}^{AHE}$ are excluded. Furthermore, <AHE> rapidly decreases with increasing SIO thickness and since the LCO/STO/SIO trilayer does not display any AHE throughout the complete temperature range, the measurements document proximity induced ferromagnetism in the SIO layer by LCO.

Below 90 K, the AHE of LCO/SIO is rather large and dominates $\rho_{xy}$. The sign of $\rho_{xy}^{AHE}$ is always positive and opposite to that which has been recently discussed for SIO/manganite heterostructures.[34] Because of the insulating nature of LCO, the AHE unambiguously originates from the conductive SIO layer. This in turn indicates that underlying mechanism strongly relies on the related magnetic ion, Ir, which inherently shows a large SOC. The anomalous Hall effect is a direct consequence of time reversal symmetry breaking and spin orbit coupling. Depending on $\sigma_{xx}$ of the sample, intrinsic scattering which is related to the topology of the electronic band structure (through the Berry phase curvature (BC)) or extrinsic skew- and side jump scattering dominate $\sigma_{xy}^{AHE}$.[48] For LCO/SIO $\sigma_{xx} \approx 500$ $\Omega^{-1}$cm$^{-1}$ at 80 K, which puts the material into the moderately dirty metal limit, where intrinsic scattering results in the independence of $\sigma_{xy}^{AHE}$ on the transport lifetime and scattering mechanism (so to the independence of $\sigma_{xy}^{AHE}$ on $\sigma_{xx}$).[48] This is indeed the case for $T \leq 30$ K as seen in **Figure 3**c, where we show that $\sigma_{xy}^{AHE}$ remains constant as $\sigma_{xx}$ decreases. The maximum Hall angle $\sigma_{xy}^{AHE}/\sigma_{xx}$ – the figure of merit for spin- to charge-current conversion – amounts to 0.8%, a value comparable to that of iridate oxide or other heavy metals displaying strong SOC,[50] but orders of magnitude larger compared to 3$d$ metal oxides, *e. g.*, $Co_xTi_{1-x}O_{2-\delta}$.[45] This hints to nontrivial topological band properties of SIO. The band structure of three-dimensional orthorhombic (P$bnm$) SrIrO$_3$ structure displays a line node made of $J_{eff} = 1/2$ bands below the Fermi level.[51] Band crossings at the Dirac nodal ring in the *U-R-X* plane originating from the mirror reflection of the crystalline P$bnm$ symmetry, prevent a full gap opening and result in a topological metallic state.[52] Lifting the Dirac degeneracy by breaking the mirror symmetry or by epitaxial strain may lead to various topological surface states in heteroepitaxial superlattices.[53–55]

In principle, the intrinsic AHE can be calculated from the BC of the occupied Bloch bands. However, for SIO these are very sensitive to structural distortions and charge transfer, which impedes accurate calculations very much. Nevertheless, theoretical analysis indicates strong enhancement of BC due to magnetic monopoles induced at avoided band crossings in the presence of strong SOC and magnetic order.[34]

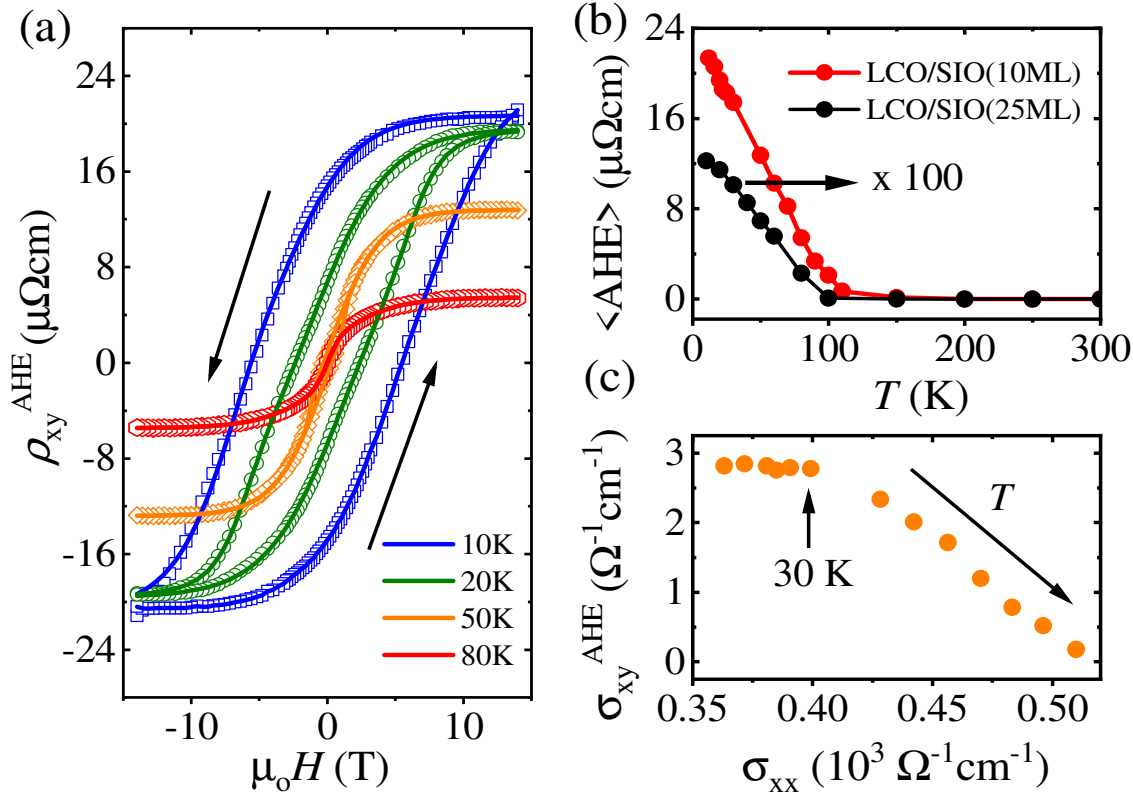

**Figure 3**. (a) $\rho_{xy}^{AHE}$ versus $\mu_0 H$ of LCO/SIO at different $T$ (symbols). More data can be found in Figure S6, supporting information. Field sweep direction is indicated. For the fits (solid lines) the field-dependence of $M$ is described by Heaviside-step functions, see text. (b) The amplitude <AHE> of $\rho_{xy}^{AHE}$ versus $T$ for LCO/SIO heterostructures with 10 ML and 25 ML (multiplied by 100) SIO thickness. (c) $\sigma_{xy}^{AHE}$ versus $\sigma_{xx}$. Data were obtained at different $T$ and $\mu_0 H = 14$ T.

Much can also be learned about this new FM state of SIO by the magnetoresistance. Below $T_C$, the LCO/SIO bilayer displays negative hysteretic contributions to the normal magnetoresistance $MR = [\rho_{xx}(H) - \rho_{xx}(0)]/\rho_{xx}(0)$, which is absent in the single- and the trilayer sample.[37] In **Figure 4**a, we show $MR$ versus $\mu_0 H$ for various temperatures between 10 and 100 K. Within that temperature range, $MR$ is well described by the sum of two contributions, namely the classical Lorentz scattering, i.e., $MR \propto H^2$,[56] and spin-flip scattering which results in a negative contribution to $MR \propto -M^2$ and which is effectively suppressed above the ferromagnetic transition temperature.[57] Even though, SIO single layers display distinct weak antilocalization (positive $MR$) at low $T$,[45] this phase coherent electronic transport is effectively suppressed by the FM order. We cannot rule out completely other minor contributions to $MR$ which however does not affect our main conclusions here. The fits to the data are shown in Figure 4a alike. Fitting parameters are listed in Table S1, see supporting information. Contributions by the spin-flip scattering are perfectly fitted by using the same Heaviside-step functional behavior for $M$ as for the AHE measurement analysis (Fig. 3a). In **Figure 4**b we show the normalized magnetization $m_{MR} = M(\mu_0 H)/M(14\ T)$ as deduced from $MR$, and compare it to that obtained from the AHE, $m_{AHE}$, for $T_C \geq T > 10$ K. The field dependence for both, $m_{MR}$ and $m_{AHE}$, is very similar, documenting the consistency of the magnetotransport with respect to the magnetic state of SIO. Discrepancies appear for $T \leq 10$ K, where no full saturation of sample magnetization can be achieved with the highest magnetic field available on our experimental set-up (14 T). Note that the magnetization at $T = 100$ K is about 6 times smaller compared to $T = 50$ K.

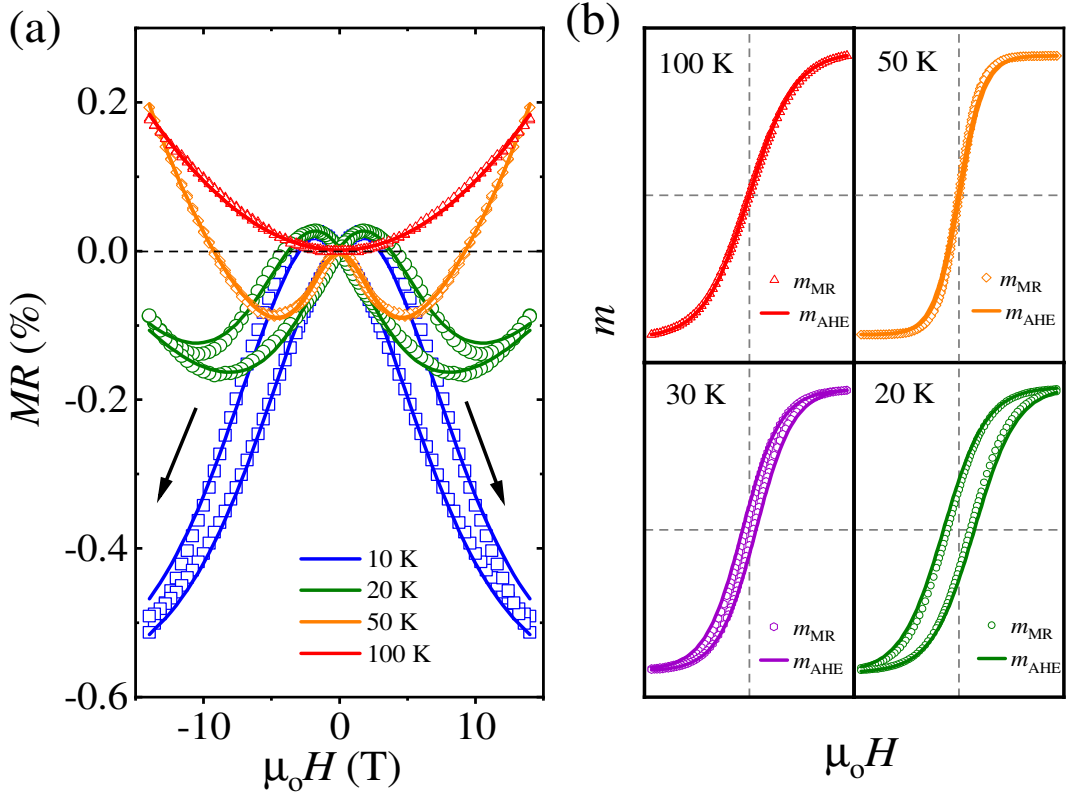

**Figure 4**. (a) *MR* versus $\mu_0 H$ of LCO/SIO at different *T* (symbols). Field sweep direction is indicated. Fits to the data, see text, are shown by solid lines. (b) The normalized magnetization $m_{MR}$ (symbols) and $m_{AHE}$ (solid lines) as deduced from fits to the *MR* and AHE, respectively, for various temperatures. Scale of each plot is the same.

The underlaying mechanism of magnetism in LCO/SIO is expected to be similar to that discussed for manganite/SIO heterostructures by Bhowal et al., *i. e.*, a combination of proximity effect and hole-doping.[58] The proximity effect which arises by the AFM coupling of the Ir atoms with the Co atoms drives the SIO layer FM to align it antiferromagnetically with the FM moments of the LCO layer. On the other side, the interfacial hybridization is inherently connected to charge transfer. In these terms, charge transfer may also affect the strength of the magnetic coupling. The electron transfer from SIO to LCO, *i. e.*, hole-doping of SIO may also support ferromagnetism in SIO.[59]

## 2.3 Magnetic anisotropy of SIO

As pointed out in the previous section, the saturation field at 10 K is extremely large (>14 T) indicating magnetic hard-axis behavior along the out-of-plane (001) direction. In the following, the magnetic anisotropy of the SIO layer is analyzed by field-sweeps along different crystallographic directions. In **Figure 5**, *MR* at 10 K is shown for out-of-plane field direction, *i. e.*, *H* parallel to the surface normal *n*, in comparison to *MR* for in-plane field direction (*H* ⊥ $n$). Here, α denotes the angle between in-plane field- and current direction. Obviously, for in-plane field directions, the coercivity is nearly vanishing and much smaller compared to *MR* for *H* parallel to *n*, which documents magnetic easy in-plane behavior. However, a closer look at the data reveals for α = 45° a somewhat larger |*MR*| compared to α = 0° or 90°. The larger |*MR*| is interpreted by a larger *M* and therefore by an in-plane <110> magnetic easy-axis.

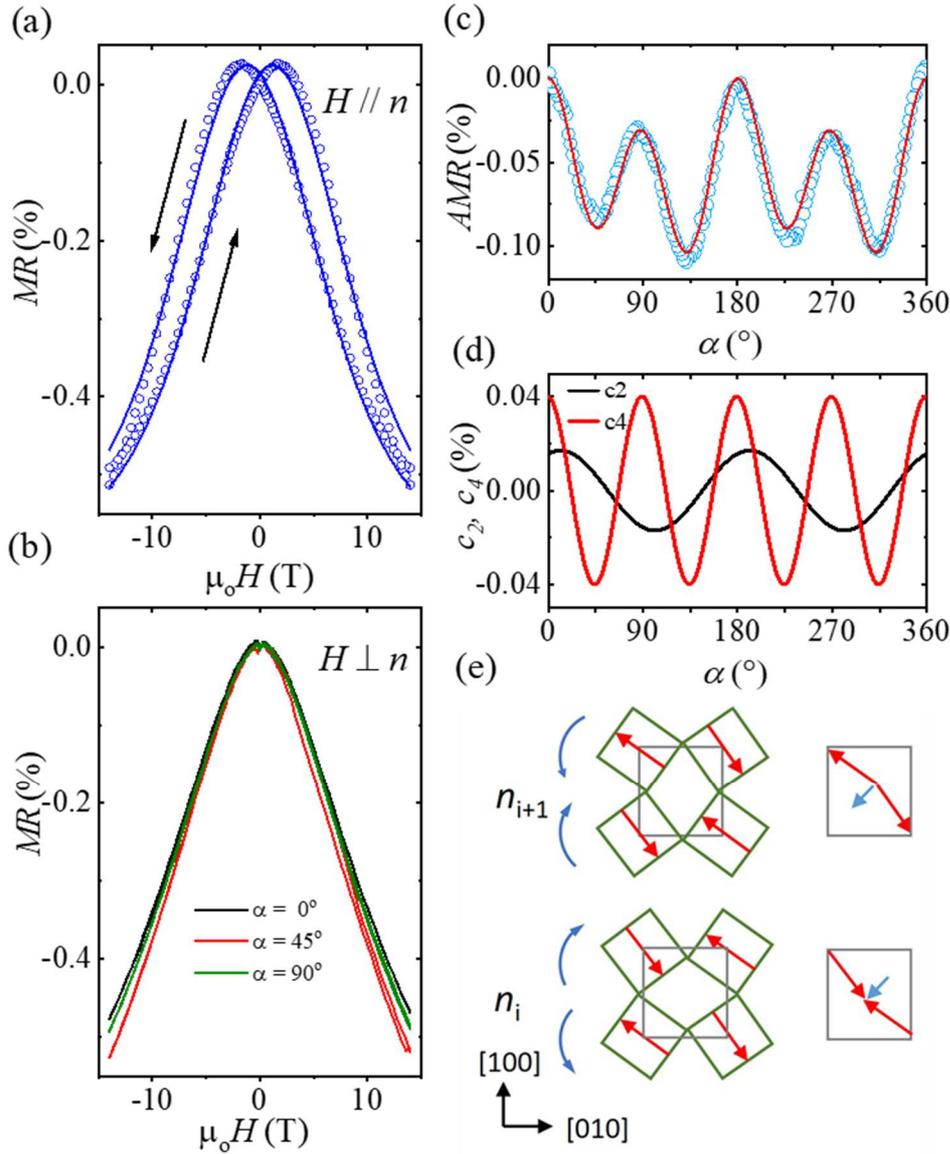

**Figure 5**. (a) *MR* of LCO/SIO versus $\mu_0 H$ at 10 K. Field direction is parallel to the surface normal *n* of the film. Field-sweep direction up (down) is indicated by arrow and green (blue) symbols. Fits to the data (see text) are shown by solid lines. (b) *MR* for in-plane field direction for different α at 10 K. (c) Anisotropic magnetoresistance at 5 K and 14T. Data are well described by $AMR(\alpha) = c_0 + c_2 \times \cos(2\alpha+\omega_2) + c_4 \times \cos(4\alpha+\omega_4)$, with the amplitudes $c_0$, $c_2$, and $c_4$, and the offset angles $\omega_2$ and $\omega_4$ (solid line). For $\alpha = \omega_2/2$ and $\omega_4/4$ magnetic field is parallel to current- and [100] direction, respectively. Fitting parameters are listed in Table S2, see supporting information. (d) The two-fold normal and four-fold magnetocrystalline component of the $AMR(\alpha)$ as deduced from the fitting results. (e) *ab*-plane view of the rotation pattern of the $IrO_6$ octahedra (green squares). The antiphase rotation along the *c*-axis is indicated by the neighbored $n_i$ and $n_{i+1}$ ML of SIO. The rotation direction is shown by arrows. The pseudospin

(red arrow) rotation is similar due to the strong pseudospin-lattice coupling in the iridates. The canting of the pseudospins results in a magnetic net moment (shown on the right by blue arrows) which due to the antiphase rotation shows FM coupling in the *ab*-plane and along the *c*-axis.

The magnetic anisotropy of the LCO/SIO bilayer is further evidenced by the anisotropic magnetoresistance $AMR(\alpha) = [\rho_{xx}(\alpha) - \rho_{xx}(0)]/\rho_{xx}(0)$, see **Figure 5**c. The normal *AMR*, the resistance difference for magnetization/field parallel to current flow and orthogonal to it, is two-fold symmetric with maxima for I parallel to magnetic field. Angle dependent Spin-Hall- or Edelstein magnetoresistance are two-fold symmetric alike however, in magnetic materials usually an order of magnitude smaller compared to the normal *AMR*. For the bilayer, a distinct $AMR(\alpha)$ appears below $T_C$ and increases with increasing field strength.[37] Beside a small two-fold symmetric contribution, $AMR(\alpha)$ is dominated by a four-fold, hence a *magnetocrystalline* anisotropy independent on current direction, with minima at $\alpha = n \times 45°$ ($n$ = 1 - 4). Here, the minima indicate the minimum of spin-flip scattering and hence indicate a dominantly parallel spin-alignment with magnetic field. For the maxima positions, there is some remaining spin-flip scattering and hence less strict spin alignment with respect to magnetic field. The minima positions are energetically the lowest for spin alignment parallel to field and present the easy-axis direction.[60] Therefore, $AMR(\alpha)$ also evidences <110> magnetic easy-axis. $AMR(\alpha)$ is well described by a two-fold normal- and a four-fold symmetric magnetocrystalline magnetoresistance: $AMR(\alpha) = c_0 + c_2 \times \cos(2\alpha + \omega_2) + c_4 \times \cos(4\alpha + \omega_4)$, where $c_0$ is angle-independent contribution to the *AMR*, and $c_2$ and $c_4$ are the amplitudes of the normal and magnetocrystalline magnetoresistance, respectively. $\omega_2$ and $\omega_4$ are the offset angles of the magnetic field direction with respect to the current- and crystallographic [100] direction, respectively.

The magnetocrystalline anisotropy of the LCO/SIO bilayer is apparently related to the rotation pattern of the $IrO_6$ octahedra. In $Sr_2IrO_4$, the $IrO_6$ octahedra are rotated in a similar way around the *c*-axis with respect to the ideal *I4/mmm* tetragonal space group.[46] Below $T_N$ = 240 K of $Sr_2IrO_4$, the pseudospins follow the rotation of the oxygen octahedra which results in a canted AFM structure in the *ab*-plane with a canting angle of about 12° from the <110>-direction.[46] In SIO, the pseudospin state of the $Ir^{4+}$ ion ($J_{eff}$ = ½) can be assumed to be similar except the magnetic ground state. The schematic of the $IrO_6$ rotation pattern for SIO of the LCO/SIO bilayer is shown in **Figure 5**d. The opposite rotation of neighbored corner-sharing octahedra around the [001]-direction favors a FM order of the net moments arising from the canted AFM order of the pseudospins of SIO with an alignment along the <110> direction. As shown before, the magnetoresistance measurements indeed advocate for a resulting magnetic in-plane <110>-easy axis. In that context, the large saturation fields for $T \leq 10$ K of the AHE and *MR* is very likely explained in terms of the strong pseudospin-lattice coupling and the freeze-out of phonons at low $T$.[16] Epitaxial strain may also enhance antiphase octahedral rotations and induce an AHE with large coercive and saturation field. However, since the octahedral distortion of SIO found in the LCO/SIO bilayer is same as for the SIO single layer, strain can be ruled out as a possible reason for this.

A crude estimation for the magnetic net moment of SIO from magnetization measurements results in about 0.1 $\mu_B$/Ir, about twice that of $Sr_2IrO_4$.[37,46] This may indicate enhanced spin-canting with respect to $Sr_2IrO_4$, possibly due to the compressive epitaxial growth of SIO on STO. These results also strongly hint to a FM ordering of the net moments within the SIO layer. Furthermore, the magnetization measurements also indicate an orientation of the SIO net moments antiparallel to those of LCO. This is similar to that found in SIO/manganite superlattices, where the Ir moments are coupled opposite to the moments of Mn.[25,26]

## 2.4 Interfacial charge transfer

The appearance of AHE in LCO/SIO below about 100 K, which is close to the $T_C$ of strained LCO films,[61] indicates that the FM state in the SIO layer is not only caused by the structural modification of SIO but also triggered by magnetic coupling to the LCO layer via interfacial charge transfer. A prerequisite for the occurrence of proximity induced magnetism and interfacial magnetic exchange is the hybridization of neighbored atomic orbitals at the interface or the formation of bonding molecular orbitals. The strength of the magnetic exchange will depend on the degree of the orbital overlap and resulting charge transfer. Because of charge neutrality, a decrease of electrical charge by $\Delta n$ on the one side must lead to an increase by $\Delta n$ on the other side.

To identify possible charge transfer at the LCO/SIO interface, we investigated the valence state of Co with respect to the distance $d$ from the interface by atomically resolved electron energy loss spectrum (EELS) imaging in scanning transmission electron microscopy (STEM) mode.[37] Co $L_3$ and $L_2$ signals were integrated to construct the Co map shown in **Figure 6**a (right) together with the high-angle annular dark-field (HAADF) STEM image (left). Co atomic columns can be clearly distinguished from the Co map and the LCO/SIO interface is atomically sharp. To quantitatively analyze the Co valence states with respect to $d$ from the interface, EELS spectra were integrated from the Co columns directly ($d = 0$) and close ($d = 1$-$3$ ML) to the interface, as indicated by the labeled boxes in the STEM image. The Co $L_3$ absorption peak clearly shifts to lower energy (**Figure S9**b) and shows increased $L_3/L_2$ ratio towards the interface. The $L_3/L_2$ intensity ratio abruptly drops at a distance of 2 ML from the interface, see **Figure 6**b. The peak ratio reflects the $3d$ occupancy and evidences a distinct lowering of the valence state of the Co ion from 3+ towards 2+ at the interface ($d = 0$). The measurements indicate confined electron transfer from Ir to Co which may hint to the formation of Ir-Co bonding molecular orbitals, akin to recent observation made in SIO/manganite heterostructures.[29]

The rather large energy of the Ir $L_3/L_2$ abortion edge (> 11.2 keV) did not allow similar EELS analysis within a TEM for Ir atoms close to the interface. Verification of a change of the Ir valence state by for example x-ray absorption spectroscopy at the Ir-L edge may also be very ambiguous.[27] In contrast, due to the rather low charge carrier concentration of SIO, much larger sensitivity and reliability towards $\Delta n$ is achieved by electrical transport.

For that reason, we have estimated the electron loss of the Ir atoms by Hall measurements. At $T = 150$ K, well above $T_C$, the Hall resistivity ($\rho_{xy} = \rho_{xy}^{OHE}$) for the LCO/SIO bilayer and the LCO/STO/SIO trilayer is strictly linear, see **Figure 6**c. The slope of $\rho_{xy}$ versus $\mu_0 H$ is negative and larger for the LCO/SIO bilayer indicating a larger $R^O$ and hence lower electron concentration. For the bi- and trilayer we deduced $n_e = 1 \times 10^{21}$ cm$^{-3}$ and $1.9 \times 10^{21}$ cm$^{-3}$. The difference $\Delta n_e = 9 \times 10^{20}$ cm$^{-3}$ (0.05 electrons per unit cell) amounts to 0.5 electrons per Ir if charge carrier depletion is confined only to the first Ir layer. This would result in a nominal valence state of Ir$^{4.5+}$ at the interface which is well comparable to the charge transfer observed in SIO/manganite heterostructures.[62] So, both experiments consistently demonstrate an electron accumulation in LCO and a depletion in SIO at the interface.

Predictions for the band alignment and charge transfer in complex oxide interfaces haven been proposed recently.[63] The alignment of oxygen states at the interface generally yields a mismatch $\Delta$ in the heterostructure`s Fermi energy driving a charge transfer. For LCO/SIO $\Delta$ amounts to about 2 eV which is very likely the reason for the charge transfer from SIO to LCO. The alignment of the Ir$5d$ and the Co$3d$ bands for such a charge transfer is sketched in **Figure 6**d.

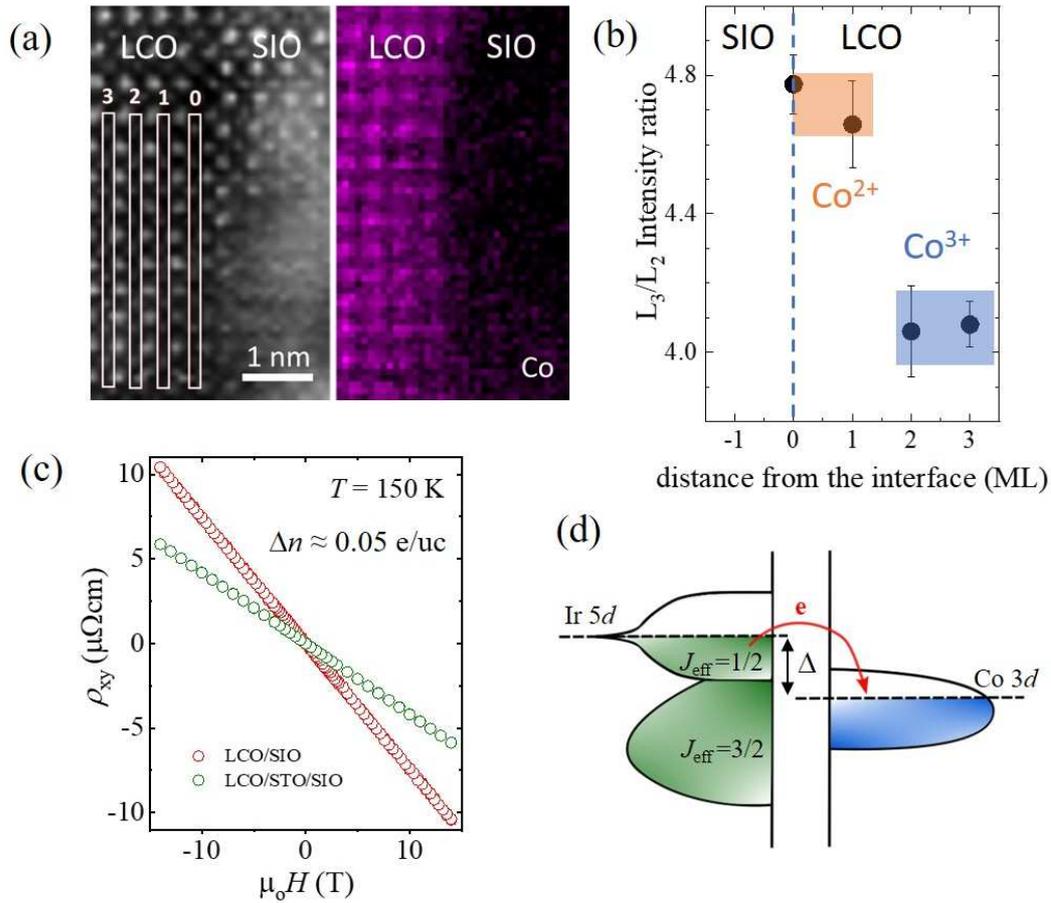

**Figure 6**. (a) HAADF-STEM image (left) and Co map from STEM-EELS spectrum imaging (right) for LCO/SIO interface. The first 4 Co rows distant from the interface are labeled and indicated by boxes. (b) The Co $L_3/L_2$ intensity ratio as a function of the distance $d$ from the interface. The Co valence state as deduced from the ratio is indicated. (c) The Hall resistivity at $T > T_C$ for the LCO/SIO bilayer and the LCO/STO/SIO trilayer. The different slopes indicate a decrease of the charge carrier concentration in LCO/SIO. (d) Schematic of the Ir5$d$ and Co3$d$ band alignment at the LCO/SIO interface. The alignment of oxygen states at the interface generally yield a mismatch in the heterostructure`s Fermi energy driving a charge transfer from SIO to LCO.

## 3. Conclusion

In conclusion, we have presented a detailed study on SIO thin film heterostructures and document a proximity induced FM state in SIO via interfacial charge transfer by using a FM insulator in contact with SIO. In contrast to previous reports on SIO heterostructures, we were able to probe electrical transport selectively only on a single SIO layer and to observe directly the FM properties of SIO. The deposition of 10 ML of SIO on STO results in a tetragonal structure with *I4/mcm* symmetry in contrast to the orthorhombic *Pbnm* symmetry of bulk material. The epitaxial growth induces antiphase octahedral rotations around the *c*-axis with a rotation pattern $a^0a^0c^-$, which enables a canted AFM state of the Ir pseudospins. However, distinct FM order of the magnetic net moments only appears in LCO/SIO heterostructures, where SIO is in direct contact with LCO. The strong positive AHE and four-fold symmetric

*AMR* evidence a FM state with $T_C \approx 100$ K, a magnetic net moment of about 0.1 $\mu_B$/Ir and a <110> in-plane magnetic easy-axis. In contrast to manganite/SIO heterostructures, the LCO/SIO bilayers display positive AHE throughout the measured temperature range. For a better understanding of this behavior a more profound theoretical elaboration including interfacial charge transfer, epitaxial strain and Rashba effect is needed. Furthermore, the AHE displays unusual large saturation field at low *T* which documents the strong pseudospin-lattice coupling in iridates. In comparison to 3*d* metal oxides the Hall angle of 0.8% is rather large hinting to nontrivial topological band properties of SIO. The results show how efficiently the electronic structure of SIO can be manipulated by structural modifications at the interface with 3*d* TMOs and may provide a new route to functionalize 5*d* metal oxide materials.

## 4. Experimental Section/Methods

*Thin film and sample preparation:*
Thin film preparation of epitaxial perovskite $SrIrO_3$ (SIO), $LaCoO_3$ (LCO), and $SrTiO_3$ (STO) was carried out by pulsed laser deposition (PLD) from stoichiometric targets on (001) oriented $TiO_2$-terminated STO substrates.[64,65] The deposition of SIO and STO was carried out at substrate temperature $T_s = 600$°C, oxygen partial pressure $p(O_2) = 0.1$ mbar, and laser fluence of $F \approx 1$ J/cm$^2$, whereas LCO was deposited at $T_s = 650$°C, $p(O_2) = 0.3$ mbar and $F \approx 2$ J/cm$^2$. The film growth was monitored by in-situ high pressure reflection high energy electron diffraction (RHEED), documenting a layer-by-layer growth mode for all the films and allowing thickness control of the deposition on the scale of one monolayer (ML).[37] For transport measurements, microbridges with a length of 200 μm and a width of 40 μm were patterned into a 6 point Hall bar geometry by ultraviolet photolithography. Contacts to the SIO layer were prepared by ultrasonic Al-wire bonding.

*High resolution electron microscopy analysis:*
The cross-sectional TEM specimen of the SIO/LCO bilayer thin film was prepared by focused ion beam (FIB) (Strata dual beam, FEI Company). The interface structure at atomic resolution was imaged by a Thermo Fisher Scientific Themis Z electron microscope, operating at 300kV and equipped with both image and probe correctors. The atomically resolved EELS spectrum imaging was performed on the same microscope in STEM mode with Gatan Continuum K3 HR image filter.

*Transport and magnetization measurements:*
The electronic transport of the samples was characterized by four-point resistance measurements on microbridges using a standard physical property measurement system equipped with a 14 T superconducting solenoid magnet and a sample holder allowing for axial sample rotation. The magnetic properties of the samples were studied by superconducting quantum interference device (SQUID) magnetometry.

**Supporting Information**
Supporting Information is available from the Wiley Online Library or from the author.

**Acknowledgement**

A. K. J. acknowledges financial support from the European Union's Framework Programme for Research and Innovation, Horizon 2020, under the Marie Skłodowska-Curie grant agreement No. 847471 (QUSTEC). We are grateful to R. Thelen and the Karlsruhe Nano-Micro Facility (KNMF) for technical support. We thank S. Mukherjee and M. Opel for fruitful discussion.

References:

[1] G. Cao, L. de Long, *Frontiers of 4D- And 5d-Transition Metal Oxides*, World Scientific Publishing Company, **2013**.
[2] D. Pesin, L. Balents, *Nature Physics* **2010**, *6*, 376.
[3] Z. Xiao, Z. Haijun, W. Jing, F. Claudia, Z. Shou-Cheng, *Science* **2012**, *335*, 1464.
[4] D. Xiao, W. Zhu, Y. Ran, N. Nagaosa, S. Okamoto, *Nature Communications* **2011**, *2*, 596.
[5] X. Wan, A. Vishwanath, S. Y. Savrasov, *Physical Review Letters* **2012**, *108*, 146601.
[6] A. Rüegg, G. A. Fiete, *Physical Review Letters* **2012**, *108*, 46401.
[7] B. J. Kim, H. Jin, S. J. Moon, J.-Y. Kim, B.-G. Park, C. S. Leem, J. Yu, T. W. Noh, C. Kim, S.-J. Oh, J.-H. Park, V. Durairaj, G. Cao, E. Rotenberg, *Physical Review Letters* **2008**, *101*, 76402.
[8] B. J. Kim, H. Ohsumi, T. Komesu, S. Sakai, T. Morita, H. Takagi, T. Arima, *Science* **2009**, *323*, 1329.
[9] G. Jackeli, G. Khaliullin, *Physical Review Letters* **2009**, *102*, 17205.
[10] S. Boseggia, R. Springell, H. C. Walker, H. M. Rønnow, Ch. Rüegg, H. Okabe, M. Isobe, R. S. Perry, S. P. Collins, D. F. McMorrow, *Physical Review Letters* **2013**, *110*, 117207.
[11] S. Fujiyama, K. Ohashi, H. Ohsumi, K. Sugimoto, T. Takayama, T. Komesu, M. Takata, T. Arima, H. Takagi, *Physical Review B* **2012**, *86*, 174414.
[12] M. A. Zeb, H.-Y. Kee, *Physical Review B* **2012**, *86*, 85149.
[13] Y. F. Nie, P. D. C. King, C. H. Kim, M. Uchida, H. I. Wei, B. D. Faeth, J. P. Ruf, J. P. C. Ruff, L. Xie, X. Pan, C. J. Fennie, D. G. Schlom, K. M. Shen, *Physical Review Letters* **2015**, *114*, 16401.
[14] L. Zhang, Q. Liang, Y. Xiong, B. Zhang, L. Gao, H. Li, Y. B. Chen, J. Zhou, S.-T. Zhang, Z.-B. Gu, S. Yao, Z. Wang, Y. Lin, Y.-F. Chen, *Physical Review B* **2015**, *91*, 35110.
[15] D. J. Groenendijk, N. Manca, G. Mattoni, L. Kootstra, S. Gariglio, Y. Huang, E. van Heumen, A. D. Caviglia, *Applied Physics Letters* **2016**, *109*, 041906.
[16] J. Porras, J. Bertinshaw, H. Liu, G. Khaliullin, N. H. Sung, J.-W. Kim, S. Francoual, P. Steffens, G. Deng, M. M. Sala, A. Efimenko, A. Said, D. Casa, X. Huang, T. Gog, J. Kim, B. Keimer, B. J. Kim, *Physical Review B* **2019**, *99*, 85125.
[17] A. M. Glazer, *Acta Crystallographica Section B* **1972**, *28*, 3384.
[18] P. E. R. Blanchard, E. Reynolds, B. J. Kennedy, J. A. Kimpton, M. Avdeev, A. A. Belik, *Physical Review B* **2014**, *89*, 214106.
[19] P. Schütz, D. di Sante, L. Dudy, J. Gabel, M. Stübinger, M. Kamp, Y. Huang, M. Capone, M.-A. Husanu, V. N. Strocov, G. Sangiovanni, M. Sing, R. Claessen, *Physical Review Letters* **2017**, *119*, 256404.
[20] R. Chaurasia, K. Asokan, K. Kumar, A. K. Pramanik, *Physical Review B* **2021**, *103*, 64418.

[21] L. Hao, D. Meyers, C. Frederick, G. Fabbris, J. Yang, N. Traynor, L. Horak, D. Kriegner, Y. Choi, J.-W. Kim, D. Haskel, P. J. Ryan, M. P. M. Dean, J. Liu, *Physical Review Letters* **2017**, *119*, 27204.

[22] L. Hao, D. Meyers, H. Suwa, J. Yang, C. Frederick, T. R. Dasa, G. Fabbris, L. Horak, D. Kriegner, Y. Choi, J.-W. Kim, D. Haskel, P. J. Ryan, H. Xu, C. D. Batista, M. P. M. Dean, J. Liu, *Nature Physics* **2018**, *14*, 806.

[23] J. Matsuno, K. Ihara, S. Yamamura, H. Wadati, K. Ishii, V. V. Shankar, H.-Y. Kee, H. Takagi, *Physical Review Letters* **2015**, *114*, 247209.

[24] J. Yang, L. Hao, D. Meyers, T. Dasa, L. Xu, L. Horak, P. Shafer, E. Arenholz, G. Fabbris, Y. Choi, D. Haskel, J. Karapetrova, J.-W. Kim, P. J. Ryan, H. Xu, C. D. Batista, M. P. M. Dean, J. Liu, *Physical Review Letters* **2020**, *124*, 177601.

[25] M. Jobu, O. Naoki, Y. Kenji, K. Fumitaka, K. Wataru, N. Naoto, T. Yoshinori, K. Masashi, *Science Advances* **2021**, *2*, e1600304.

[26] D. Yi, J. Liu, S.-L. Hsu, L. Zhang, Y. Choi, J.-W. Kim, Z. Chen, J. D. Clarkson, C. R. Serrao, E. Arenholz, P. J. Ryan, H. Xu, R. J. Birgeneau, R. Ramesh, *Proceedings of the National Academy of Sciences* **2016**, *113*, 6397.

[27] J. Nichols, X. Gao, S. Lee, T. L. Meyer, J. W. Freeland, V. Lauter, D. Yi, J. Liu, D. Haskel, J. R. Petrie, E.-J. Guo, A. Herklotz, D. Lee, T. Z. Ward, G. Eres, M. R. Fitzsimmons, H. N. Lee, *Nature Communications* **2016**, *7*, 12721.

[28] S. Elizabeth, N. John, O. J. Mok, C. R. V, C. E. Sang, R. Ankur, S. Changhee, G. Xiang, Y. Sangmoon, F. Thomas, D. R. D, C. Yongseong, H. Daniel, F. J. W, O. Satoshi, B. Matthew, L. H. Nyung, *Science Advances* **2021**, *6*, eaaz3902.

[29] Y. Zhang, Y. Z. Luo, L. Wu, M. Suzuki, Q. Zhang, Y. Hirata, K. Yamagami, K. Takubo, K. Ikeda, K. Yamamoto, A. Yasui, N. Kawamura, C. Lin, K. Koshiishi, X. Liu, J. Zhang, Y. Hotta, X. R. Wang, A. Fujimori, Y. Lin, C. Nan, L. Shen, H. Wadati, *Physical Review Research* **2020**, *2*, 33496.

[30] S. Bhowal, S. Satpathy, *Physical Review B* **2019**, *99*, 245145.

[31] D. Yi, C. L. Flint, P. P. Balakrishnan, K. Mahalingam, B. Urwin, A. Vailionis, A. T. N'Diaye, P. Shafer, E. Arenholz, Y. Choi, K. H. Stone, J.-H. Chu, B. M. Howe, J. Liu, I. R. Fisher, Y. Suzuki, *Physical Review Letters* **2017**, *119*, 77201.

[32] T. S. Suraj, G. J. Omar, H. Jani, M. M. Juvaid, S. Hooda, A. Chaudhuri, A. Rusydi, K. Sethupathi, T. Venkatesan, A. Ariando, M. S. R. Rao, *Physical Review B* **2020**, *102*, 125145.

[33] G. A. Ovsyannikov, T. A. Shaikhulov, K. L. Stankevich, Yu. Khaydukov, N. v Andreev, *Physical Review B* **2020**, *102*, 144401.

[34] M.-W. Yoo, J. Tornos, A. Sander, L.-F. Lin, N. Mohanta, A. Peralta, D. Sanchez-Manzano, F. Gallego, D. Haskel, J. W. Freeland, D. J. Keavney, Y. Choi, J. Strempfer, X. Wang, M. Cabero, H. B. Vasili, M. Valvidares, G. Sanchez-Santolino, J. M. Gonzalez-Calbet, A. Rivera, C. Leon, S. Rosenkranz, M. Bibes, A. Barthelemy, A. Anane, E. Dagotto, S. Okamoto, S. G. E. te Velthuis, J. Santamaria, J. E. Villegas, *Nature Communications* **2021**, *12*, 3283.

[35] P. W. Anderson, H. Hasegawa, *Physical Review* **1955**, *100*, 675.

[36] A. Maignan, V. Caignaert, B. Raveau, D. Khomskii, G. Sawatzky, *Physical Review Letters* **2004**, *93*, 26401.

[37] **Supporting Information**

[38] A. K. Jaiswal, R. Schneider, R. Singh, D. Fuchs, *Applied Physics Letters* **2019**, *115*, 031904.

[39] G. Thornton, B. C. Tofield, A. W. Hewat, *Journal of Solid State Chemistry* **1986**, *61*, 301.

[40] W. Guo, D. X. Ji, Z. B. Gu, J. Zhou, Y. F. Nie, X. Q. Pan, *Physical Review B* **2020**, *101*, 85101.

[41] L. Qiao, J. H. Jang, D. J. Singh, Z. Gai, H. Xiao, A. Mehta, R. K. Vasudevan, A. Tselev, Z. Feng, H. Zhou, S. Li, W. Prellier, X. Zu, Z. Liu, A. Borisevich, A. P. Baddorf, M. D. Biegalski, *Nano Letters* **2015**, *15*, 4677.
[42] A. Biswas, K.-S. Kim, Y. H. Jeong, *Journal of Applied Physics* **2014**, *116*, 213704.
[43] L. Zhang, Y. B. Chen, B. Zhang, J. Zhou, S. Zhang, Z. Gu, S. Yao, Y. Chen, *Journal of the Physical Society of Japan* **2014**, *83*, 054707.
[44] F.-X. Wu, J. Zhou, L. Y. Zhang, Y. B. Chen, S.-T. Zhang, Z.-B. Gu, S.-H. Yao, Y.-F. Chen, *Journal of Physics: Condensed Matter* **2013**, *25*, 125604.
[45] D. J. Groenendijk, C. Autieri, J. Girovsky, M. C. Martinez-Velarte, N. Manca, G. Mattoni, A. M. R. V. L. Monteiro, N. Gauquelin, J. Verbeeck, A. F. Otte, M. Gabay, S. Picozzi, A. D. Caviglia, *Physical Review Letters* **2017**, *119*, 256403.
[46] F. Ye, S. Chi, B. C. Chakoumakos, J. A. Fernandez-Baca, T. Qi, G. Cao, *Physical Review B* **2013**, *87*, 140406.
[47] J. Kondo, *Progress of Theoretical Physics* **1964**, *32*, 37.
[48] N. Nagaosa, J. Sinova, S. Onoda, A. H. MacDonald, N. P. Ong, *Reviews of Modern Physics* **2010**, *82*, 1539.
[49] D. Fuchs, P. Schweiss, P. Adelmann, T. Schwarz, R. Schneider, *Physical Review B* **2005**, *72*, 14466.
[50] K. Fujiwara, Y. Fukuma, J. Matsuno, H. Idzuchi, Y. Niimi, Y. Otani, H. Takagi, *Nature Communications* **2013**, *4*, 2893.
[51] J.-M. Carter, V. V. Shankar, M. A. Zeb, H.-Y. Kee, *Physical Review B* **2012**, *85*, 115105.
[52] M. A. Zeb, H.-Y. Kee, *Physical Review B* **2012**, *86*, 85149.
[53] J. Liu, D. Kriegner, L. Horak, D. Puggioni, C. Rayan Serrao, R. Chen, D. Yi, C. Frontera, V. Holy, A. Vishwanath, J. M. Rondinelli, X. Marti, R. Ramesh, *Physical Review B* **2016**, *93*, 85118.
[54] Y. Chen, H.-Y. Kee, *Physical Review B* **2014**, *90*, 195145.
[55] Y. Chen, Y.-M. Lu, H.-Y. Kee, *Nature Communications* **2015**, *6*, 6593.
[56] D. Feng, G. Jin, *Introduction to Condensed Matter Physics*, World Scientific Publishing Company, **2005**.
[57] K. Yosida, *Physical Review* **1957**, *107*, 396.
[58] S. Bhowal, S. Satpathy, *npj Computational Materials* **2019**, *5*, 61.
[59] Y. Nagaoka, *Physical Review* **1966**, *147*, 392.
[60] N. Lee, E. Ko, H. Y. Choi, Y. J. Hong, M. Nauman, W. Kang, H. J. Choi, Y. J. Choi, Y. Jo, *Advanced Materials* **2018**, *30*, 1805564.
[61] D. Fuchs, E. Arac, C. Pinta, S. Schuppler, R. Schneider, H. v. Löhneysen, *Physical Review B* **2008**, *77*, 14434.
[62] S. Okamoto, J. Nichols, C. Sohn, S. Y. Kim, T. W. Noh, H. N. Lee, *Nano Letters* **2017**, *17*, 2126.
[63] Z. Zhong, P. Hansmann, *Physical Review X* **2017**, *7*, 11023.

[64] K. R. Kleindienst, K. Wolff, J. Schubert, R. Schneider, D. Fuchs, *Physical Review B* **2018**, *98*, 115113.
[65] D. Fuchs, C. Pinta, T. Schwarz, P. Schweiss, P. Nagel, S. Schuppler, R. Schneider, M. Merz, G. Roth, H. v. Löhneysen, *Physical Review B* **2007**, *75*, 144402.